\documentclass{ieeeaccess}
\usepackage{cite}
\usepackage{amsmath,amssymb,amsfonts}
\usepackage{algorithmic}
\usepackage{graphicx}
\usepackage{textcomp}

\usepackage{diagbox}
\usepackage{subfiles}
\usepackage{style/customize}
\usepackage{physics}
\usepackage{threeparttable}
\usepackage{multirow}

\newcommand\figref{Fig.~\ref}
\newcommand\tabref{Table~\ref}
\newcommand\secref{Section~\ref}

\def\BibTeX{{\rm B\kern-.05em{\sc i\kern-.025em b}\kern-.08em
    T\kern-.1667em\lower.7ex\hbox{E}\kern-.125emX}}
\begin{document}
\doi{??.????/???.????.???}

\title{C3-VQA: Cryogenic Counter-Based Co-Processor for Variational Quantum Algorithms}
\author{
\uppercase{Yosuke Ueno}\authorrefmark{1}, 
\uppercase{Satoshi Imamura}\authorrefmark{2},
\uppercase{Yuna Tomida}\authorrefmark{3}, 
\uppercase{Teruo Tanimoto}\authorrefmark{4}, 
\uppercase{Masamitsu Tanaka}\authorrefmark{5}, 
\uppercase{Yutaka Tabuchi}\authorrefmark{1}, 
\uppercase{Koji Inoue}\authorrefmark{4}, and 
\uppercase{Hiroshi Nakamura}\authorrefmark{3} 
}
\address[1]{RIKEN Center for Quantum Computing, Wako, Saitama 351-0198, Japan}
\address[2]{Fujitsu Limited, Kawasaki, Kanagawa 211-8588, Japan}
\address[3]{Graduate School of Information Science and Technology, The University of Tokyo, Bunkyo, Tokyo 113-8656, Japan}
\address[4]{Faculty of Information Science and Electrical Engineering, Kyushu University, Nishi-ku, Fukuoka 819-0395, Japan}
\address[5]{Graduate School of Engineering, Nagoya University, Nagoya, Aichi 464-8603, Japan}
\tfootnote{This work was partly supported by JST Moonshot R\&D Grant Number JPMJMS2067, JSPS KAKENHI Grant Numbers JP22H05000, JP22K17868, RIKEN Special Postdoctoral Researcher Program.}

\markboth
{Ueno \headeretal: C3-VQA: Cryogenic Counter-Based Co-Processor for Variational Quantum Algorithms}
{Ueno \headeretal: C3-VQA: Cryogenic Counter-Based Co-Processor for Variational Quantum Algorithms}

\corresp{Corresponding author: Yosuke Ueno (email: yosuke.ueno@riken.jp)}

\begin{abstract}

Cryogenic quantum computers play a leading role in demonstrating quantum advantage.
Given the severe constraints on the cooling capacity in cryogenic environments, thermal design is crucial for the scalability of these computers. 
The sources of heat dissipation include passive inflow via inter-temperature wires and the power consumption of components located in the cryostat, such as wire amplifiers and quantum-classical interfaces. 
Thus, a critical challenge is to reduce the number of wires by reducing the required inter-temperature bandwidth while maintaining minimal additional power consumption in the cryostat.

One solution to address this challenge is near-data processing using ultra-low-power computational logic within the cryostat. 
Based on the workload analysis and domain-specific system design focused on Variational Quantum Algorithms~(VQAs), we propose the Cryogenic Counter-based Co-processor for VQAs (C3-VQA) to enhance the design scalability of cryogenic quantum computers under the thermal constraint.
The C3-VQA utilizes single-flux-quantum logic, which is an ultra-low-power superconducting digital circuit that operates at the 4~K environment.
The C3-VQA precomputes a part of the expectation value calculations for VQAs and buffers intermediate values using simple bit operation units and counters in the cryostat, thereby reducing the required inter-temperature bandwidth with small additional power consumption.
Consequently, the C3-VQA reduces the number of wires, leading to a reduction in the total heat dissipation in the cryostat.
Our evaluation shows that the C3-VQA reduces the total heat dissipation at the 4~K stage by 30\% and 81\% under sequential-shot and parallel-shot execution scenarios, respectively.
Furthermore, a case study in quantum chemistry shows that the C3-VQA reduces total heat dissipation by 87\% with a 10,000-qubit system.

\end{abstract}


\titlepgskip=-15pt

\maketitle

\section{Introduction}
\label{sec:introduction}
Quantum computers that exploit quantum mechanical effects are expected to offer advantages in solving classically intractable problems, such as quantum chemistry, combinatorial optimization, machine learning, and fundamental algorithms like factorization and database search.
Scaling up quantum computers is essential as demonstrating the quantum advantage requires solving large-scale problems.
To achieve quantum advantage with Noisy Intermediate-Scale Quantum~(NISQ) computers, the focus of development is shifting from merely increasing the number of qubits to scaling up the entire system.
Superconducting quantum computers, which operate at extremely low temperatures, specifically around 20~mK, are currently the most promising.
To scale up superconducting quantum computers, it is necessary not only to increase the number of qubits in Quantum Processing Units~(QPUs) but also to address the associated system engineering challenges.
In particular, Quantum-Classical Interfaces~(QCIs), which control and measure qubits and manage communication between components must be scalable, taking the cryogenic environment into account.
Notable efforts include integrating classical information processing circuits within the cryostat, such as DigiQ~\cite{jokar2022digiq}, XQsim~\cite{byun2022xqsim}, and QIsim~\cite{min2023qisim}.

\Figure[t!](topskip=0pt, botskip=0pt, midskip=-2mm){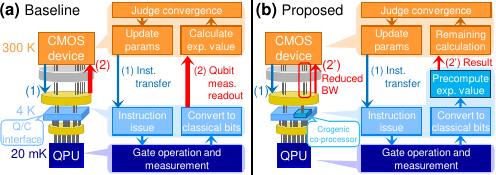}
{(a) Baseline and (b) proposed system designs.\label{fig:cryo-vqe}}


\figref{fig:cryo-vqe}\,(a) shows our baseline system design, where a cryogenic QCI is installed at the 4~K stage to address the scalability issues of a QPU and QCI.
Although the cryogenic QCI helps reduce the number of wires, heat dissipation remains a critical issue. 
This is due to passive heat inflow introduced by wiring across different stages and the power consumption of the QCI and wire peripherals, such as amplifiers, which generates heat in the cryogenic stages.
System design exploration involving a cryogenic QCI is required to build scalable cryogenic quantum computers.

However, such exploration involving a cryogenic QCI is not yet fully mature due to a lack of consideration for target workloads from a computer systems perspective.
Appropriate domain-specific application focus and system designs must be considered to achieve quantum advantage at an early stage.

The objective of our research is to establish an efficient and scalable quantum system architecture under thermal constraints.
To realize such a system, we leverage ultra-low-temperature computational techniques along with domain-specific architectural designs.
By analyzing workloads and features throughout the entire execution process, we aim to mitigate engineering challenges related to NISQ scalability by using ultra-low-power classical information processing techniques at the 4~K stage with single-flux-quantum (SFQ) circuits.

We focus on Variational Quantum Algorithms~(VQAs), which are highly regarded in the field of NISQ computing, to design a domain-specific system. 
Specifically, we consider prominent algorithms such as Quantum Approximate Optimization Algorithm~(QAOA), Variational Quantum Eigensolver~(VQE), and Quantum Machine Learning~(QML) for neural networks.
VQAs obtain expectation values through quantum state sampling while optimizing classical parameters.
This process involves executing numerous shots of the same quantum circuit and collecting qubit measurement results.
Current NISQ computer designs treat each sample acquisition as an independent task, transferring quantum instructions and measurement results in and out of the cryostat for every sample.
Our workload characterization reveals that a) the bandwidth necessary to hide communication latency grows with the number of employed qubits, and b) measurement readout dominates the bandwidth.
Additionally, we reveal that c) a simple co-processor in the cryostat can reduce the required bandwidth between the out-of-cryostat environment and the 4~K stage.

In this work, we propose a cryogenic counter-based co-processor for VQAs (\textit{C3-VQA}) using SFQ circuits in the 4 K stage of the cryostat by employing traditional architectural approaches that confine frequent repetitive operations near the processing unit to avoid costly communication.
\figref{fig:cryo-vqe}\,(b) illustrates our proposed system design.
The cryogenic counter-based co-processor precomputes measurement results for expectation value calculations and buffers the data to reduce the required bandwidth in and out of the cryostat. 
Moreover, the C3-VQA architecture is sufficiently power-efficient to operate in a cryogenic environment.
As a result, the C3-VQA successfully reduces the number of inter-temperature wires required in the system with small additional power consumption, leading to a reduction in total heat dissipation within the cryogenic environment, including the passive heat inflow from wires and the power consumption of wire peripherals and the C3-VQA itself.

We evaluate two VQA execution models: the \textit{sequential-shot (SS)} and \textit{parallel-shot (PS)} execution models.
The SS execution sequentially collects the numerous samples necessary for expectation value calculations.
In contrast, the PS execution collects samples in parallel by fully employing a large number of qubits.

We emphasize the importance of the PS scenario in the following discussion.
NISQ machines with approximately 100 qubits have the capability to perform computations at a scale beyond classical brute-force methods~\cite{Kim2023}.
However, the number of qubits employed in NISQ computations remains limited to tens or hundreds due to factors such as error rates~\cite{ichikawa2023comprehensive}.
During the transitional from NISQ to Fault-Tolerant Quantum Computing (FTQC), the number of qubits deployed in quantum computers is expected to increase to tens or hundreds of thousands to achieve quantum acceleration~\cite{yoshioka2024hunting,akahoshi2024partially}.
Therefore, the gap between the number of deployed and employed qubits in NISQ computation becomes increasingly pronounced.
The PS scenario, which involves executing a NISQ algorithm in parallel across a large number of qubits, is essential for bridging this gap and ensuring the effective utilization of all deployed qubits.
In this work, we discuss how our VQA-specialized SFQ co-processor is well-suited for PS execution.

Our evaluation shows that C3-VQA reduces the total heat dissipation at the 4~K stage by 30\% and 81\% under the SS and PS scenarios, respectively.
In addition, a case study on quantum chemistry problems shows that total heat dissipation is reduced by 87\% with a 10,000-qubit system.
This research provides the insight that traditional architectural techniques developed for higher performance in classical computers can also be applied to quantum computers to reduce the required bandwidth, thereby enhancing scalability.

The remainder of this paper is organized as follows:
\secref{background} describes the target algorithms and the cryogenic digital processing technology, SFQ logic.
\secref{analysis} models the inter-temperature bandwidth of VQA execution, and we describe the design of a cryogenic counter-based co-processor for VQAs in \secref{proposal}.
\secref{implementation} presents the implementation of the proposed co-processor using SFQ logic.
\secref{evaluation} evaluates the impact of our proposal on the scalability of cryogenic NISQ computers for VQAs.
\secref{sec:conclusion} summarizes the paper and discusses future prospects.

This paper extends the work presented in the IEEE Computer Architecture Letters~\cite{ueno2023inter} and the work-in-progress session of the 61st Design Automation Conference~\cite{ueno2024sfq}.
While the original papers focused on QAOA and VQE, respectively, this paper expands the scope to include QML and presents a comprehensive architecture design for VQAs.
Additionally, this paper presents a detailed design of the proposed architecture using SFQ circuits and outlines future prospects for the FTQC era.

\section{Background}
\label{background}


\subsection{Variational Quantum Algorithms~(VQA\lowercase{s})}

VQAs are promising hybrid quantum-classical algorithms designed for NISQ computers. 
\figref{fig:VQA_overview} provides an overview of VQA computation.
VQAs aim to minimize the cost function $C(\boldsymbol{\theta})$ by iteratively calculating an expectation value $f_k(\boldsymbol{\theta})$. 
This is achieved through the execution of a parameterized quantum circuit on a NISQ device and updating its parameters $\boldsymbol{\theta}$ using a classical optimizer until the expectation value converges.
The parameterized portion of a quantum circuit corresponds to a trial wave function called an \textit{ansatz}, and the iterative execution of an ansatz-based quantum circuit is a key characteristic of VQAs.
Three well-known algorithms in VQAs are VQE, QAOA, and QML. 
We briefly explain these algorithms in the subsequent part of this subsection.
\tabref{tab:notation} summarizes the notations related to VQAs used in this paper. 

\Figure[t!](topskip=0pt, botskip=0pt, midskip=-2mm)[width=0.98\columnwidth]{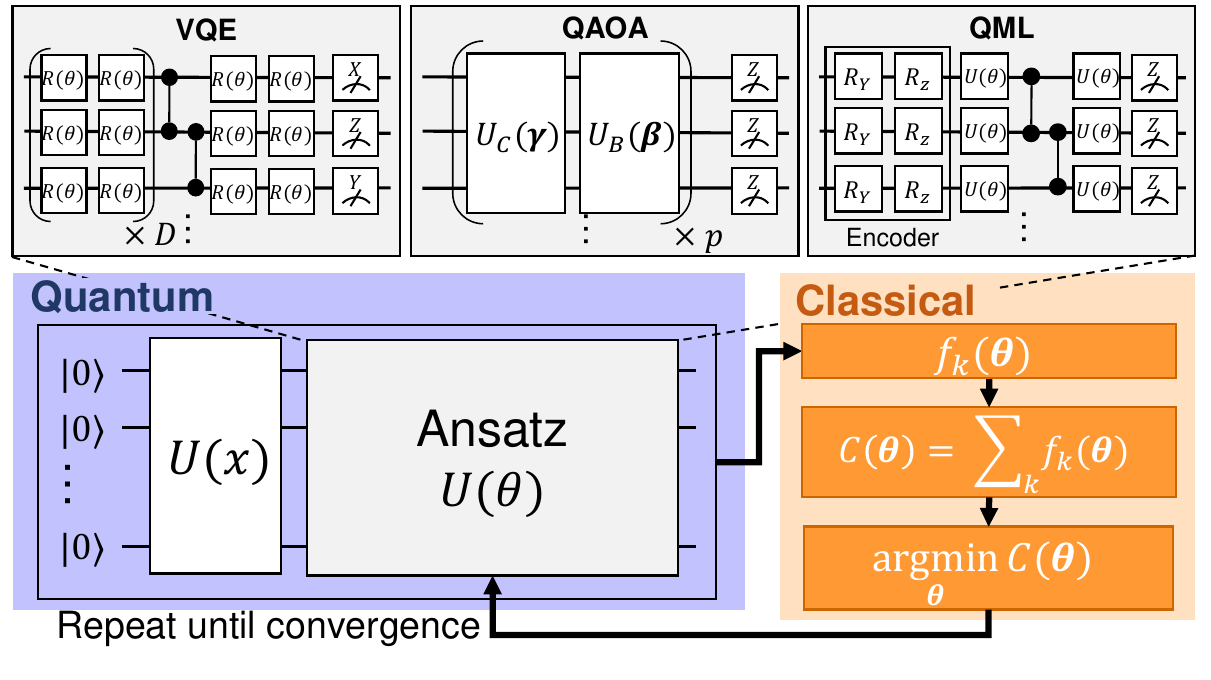}{An overview of VQAs and three example types of ansatze.\label{fig:VQA_overview}}


\subsubsection{Variational Quantum Eigensolver~(VQE)}
\label{background:vqe}

VQE is a hybrid quantum-classical algorithm that computes the ground-state energy of a Hamiltonian. 
A typical use case is calculating the ground-state energy of molecules, a fundamental task in quantum chemistry~\cite{peruzzo2014vqe}.  
VQE approximates the molecular ground-state energy by finding the optimal ansatz parameters $\boldsymbol{\theta}$ that minimize the expectation value of a Hamiltonian, $\expectation{H}$. 
The Hamiltonian of a molecule is encoded as a linear combination of Pauli terms using second quantization and a transformation method such as Jordan-Wigner~\cite{Fradkin:1989jo}, as shown in
\begin{equation}
    H = \sum_{i}w_{i}P_{i}, \label{eq:quantized_hamiltonian}
\end{equation}
where $w_{i}$ is a real-valued weight and $P_{i}$ is a Pauli string. 
Under the Jordan-Wigner transformation, the number of qubits in a quantum circuit corresponds to the number of spin orbitals of a target molecule, and the value of each qubit ($\ket{0}$ or $\ket{1}$) indicates whether an electron occupies the corresponding spin-orbital.
The expectation value of the entire Hamiltonian is calculated as
\begin{equation}
    \expectation{H} = \sum_{i}w_{i}\expectation{P_{i}}, \label{hamiltonian_expval}
\end{equation}
and the expectation value of each Pauli string, $\expectation{P_{i}}$, is calculated through $T$ times measurements (referred to as \textit{shots}) of an ansatz-based quantum circuit. 
We refer to the $T$-measurement procedure as the \textit{sampling loop} in this paper.

Since each qubit can be directly measured in the $Z$-basis, measurements in the $X$- or $Y$-basis are realized by applying basis transformation using a Hadamard or $R_x(\frac{\pi}{2})$ gate before $Z$-basis measurement, respectively.  
For instance, to calculate $\expectation{X_0Z_1Z_2Y_3}$, we prepare $\ket{\psi'(\boldsymbol{\theta})} = H_0R_x(\frac{\pi}{2})_3\ket{\psi(\boldsymbol{\theta})}$ and measure it on the $Z$-basis.
Based on the $Z$-basis measurement, $\expectation{P_{i}}$ is calculated as 
\begin{align}
\expectation{P_{i}} = \frac{C_{i, even}-C_{i, odd}}{T} = \frac{T-2C_{i, odd}}{T}, \label{eq:pauli_string_exp}
\end{align}
where $C_{i, odd}$ ($C_{i, even}$) represents the total number of measurement results in which the sum of the values of target qubits is odd (even) within  a single sampling loop.

The accuracy and computational cost of VQE strongly depend on the choice of ansatz. 
In quantum chemical calculations, \textit{unitary coupled-cluster (UCC)} ansatze are known to achieve particularly high accuracy~\cite{Ku:2019ac}. 
However, the depth and gate counts of UCC ansatze grow exponentially with the number of qubits; hence, UCC ansatze for practical molecules are too deep to be executed on current NISQ computers. 
To address this issue, \textit{hardware-efficient} ansatze~(HEA), which have much lower depth than UCC ansatze, are commonly used~\cite{kandala2017hardware}. 
As shown in the upper left in \figref{fig:VQA_overview}, the HEA has a linear depth with respect to an arbitrary $D$.

In a straightforward approach, $\expectation{H}$ is calculated by obtaining each $\expectation{P_{i}}$ and summing all the expectation values multiplied by corresponding weights, as shown in \eqref{hamiltonian_expval}.
However, this approach is time-consuming because the quantum circuit must be executed as many times as the number of Pauli strings $N_P$, which grows as $O(N^4)$~\cite{Verteletskyi:2020me}.

Group measurement (GM) is a well-known technique that groups as many Pauli strings as possible into a group $G$, enabling multiple Pauli string measurements in a single quantum circuit execution. 
In this paper, $GP$ denotes an inclusive Pauli string that covers all Pauli strings in a group $G$.
For instance, for $G = \{I_0X_1Z_2, X_0X_1I_2, I_0I_1Z_2\}$, $GP$ is $X_0X_1Z_2$.
In addition, we denote the number of groups by $N_G$ and the maximum number of Pauli strings within a group by $K$, which is called \textit{group size}.

Recent grouping methods have significantly reduced the number of quantum circuit executions (\textit{i.e.}, the number of groups $N_G$) by efficiently grouping Pauli strings~\cite{Verteletskyi:2020me, Kurita:2023pa}. 
Note that these methods reduce $N_G$ to $O(N^4)$~\cite{Verteletskyi:2020me} or $O(N)$~\cite{Kurita:2023pa}.

\begin{table}[tb]
\caption{Notations regarding VQAs used in this paper}
\vspace{-3mm}
\label{tab:notation}
\scriptsize
\centering
\scalebox{0.87}{

\begin{tabular}{|c|c|l|c|l|} \hline
Category                & Notation                 & Description                                                                             & Notation                 & Description                                   \\
\hline
\multirow{4}{*}{Common} & $N$                      & \#\tnote{*} qubits used in VQA task                                                     & $T  $                    & \# shots                                      \\
                        & $M$                      & \# qubits of a machine                                                                  & $H  $                    & Hamiltonian                                   \\
                        & $\boldsymbol{\theta}$    & Ansatz parameters                                                                       & $\ket{\psi}$             & Ansatz state                                  \\
                        & $\pz  $                  & Classical bit sequence                                                                  & $I, X, Y, Z$             & Pauli operator                                \\
\hline
\multirow{4}{*}{VQE}    & $P    $                  & Pauli string                                                                            & $N_P$                    & \# Pauli strings                              \\
                        & $w    $                  & Weight of a Pauli string                                                                & $G $                     & Group of Pauli strings                        \\
                        & $GP   $                  & Inclusive Pauli string of $G$                                                          & $N_G$                    & \# groups                                              \\
                        & $K    $                  & \begin{tabular}[c]{@{}l@{}}Group size (Max. \# \\Pauli strings in a group)\end{tabular} & $D $                     & \begin{tabular}[c]{@{}l@{}}Depth of hardware\\ efficient ansatz~\cite{kandala2017hardware}\end{tabular} \\
\hline                
\multirow{3}{*}{QAOA}   & $\pbeta$                 & Mixing Parameter                                                                        & $U_B$                    & Mixing operator                               \\
                        & $\pgam$                  & Phase paramter                                                                          & $U_C$                    & Phase operator                                \\
                        & $p    $                  & \# steps                                                                                &                          &                                               \\
\hline                        
\multirow{2}{*}{QML}    & $N_d   $                 & \# input data                                                                           & $\boldsymbol{y}$         & Class labels                                  \\
                        & $\boldsymbol{x}    $     & Training data                                                                           & $\boldsymbol{\hat{y}}$   & Output classes                                \\ \hline
                        
\end{tabular}
}
\begin{tablenotes}
  \item[*] \# in the table means ``the number of.''
\end{tablenotes}
\end{table}

\subsubsection{Quantum Approximate Quantum Algorithm~(QAOA)}
\label{background:qaoa}
QAOA is a prominent example of a VQA designed for combinatorial optimization~\cite{farhi2014qaoa}.
In QAOA, an optimization problem is formulated as a search for the ground state of the Hamiltonian of the Ising model, as shown in
\begin{align}
  H =
  - \sum_{i=1}^N h_i Z_i
  - \sum_{i \ne j} J_{ij} Z_i Z_j
  \label{eq:hamiltonian},
\end{align}
where $ Z_i $ represents the spin orientation, and $ h_i$ and $J_{ij} $ represent the magnitude of the magnetic field and the interaction between spins, respectively~\cite{inagaki2016coherent}.
The output state of the QAOA circuit is given by 
\begin{align}
  \Ket{\psi(\pgam, \pbeta)}
  = e^{ - \im \beta H_B }
  \Par{ \prod_{l=1}^{p} U_B(\beta_l) U_C(\gamma_l) }
  \Ket{+}^{\otimes N},
\end{align}
where $U_C(\gamma) = e^{ - \im \gamma H }$ is the phase operator, $U_B(\beta) = e^{ - \im \beta H_B }$ is the mixing operator, and $H_B = - \sum_{i=1}^N X_i $. 
Here, $\pgam$ and $\pbeta$ are parameters to be optimized by QAOA.

QAOA aims to minimize the expectation value of the Hamiltonian:
$
  \paramStateT{\pgam,\pbeta} H \paramState{\pgam,\pbeta}.
$.
This expectation value is approximated by the cost function $C(\pz)$, calculated as 
\begin{align}
  C(\pz) = 
  \sum_{i=1}^{N} s_{i}z_{i} +
  \sum_{i \ne j} c_{ij} \Par { z_{i} \oplus z_{j} },
  \label{eq:qaoa_cost}
\end{align}
where $ \pz \in \ZO^{N} $ is obtained from the $Z$-basis measurement of the quantum circuit. 
The real-valued coefficients $s_i$ and $c_{ij} $ correspond to $ h_i $ and $J_{ij} $ in Ising Hamiltonian, respectively.
The expectation value is calculated over $T$ shots as
\begin{align}
  \bra{\psi(\pgam, \pbeta)}H\ket{\psi(\pgam, \pbeta)} = \inv{T} \sum_{t=1}^T C(\pz_t).
  \label{eq:qaoa_sampling}
\end{align}

\subsubsection{Quantum Machine Learning~(QML)}
\label{background:qml}
QML is a type of VQA designed for performing machine learning on NISQ computers.
In particular, the quantum neural network (QNN) is a quantum deep learning algorithm inspired by classical convolutional neural networks and it can be used for image classification~\cite{Wang:2022qu}. 
In QNN, the pixels of an input image $x_k$ from a training dataset $\bm{x}$ are encoded into quantum data using rotation gates, and an ansatz is used to train the parameters $\bm{\theta}$, as shown in the upper right panel of \figref{fig:VQA_overview}. 
A classical bit sequence at the $t$-th shot, $\pz^t \in \{0, 1\}^{N}$, is obtained by measuring all qubits in the $Z$-basis, and this process is repeated $T$ times for sampling.
The probability of each class $\hat{y_k}$ is calculated from the sampling results using the softmax function $f$ as 
\begin{align}
   \hat{y_k} = f \left(\left(\sum_{t=1}^T {z_1^t}\right), ..., \left(\sum_{t=1}^T {z_N^t}\right) \right) \label{eq:QML_ff}.
\end{align}
The loss function $L_{\bm{\theta}}$ between the predicted class probabilities $\bm{\hat{y}} = \{\hat{y_1}, ...,\hat{y_{N_d}}\}$ and the actual class labels $\bm{y}$ is calculated as
\begin{align}
   L_{\bm{\theta}} = Loss (\bm{\hat{y}}, \bm{y}), \label{eq:QML_loss_function}
\end{align}
where $Loss$ is an appropriate error function, such as the cross-entropy error. 
The parameters $\bm{\theta}$ are then updated based on $L_{\bm{\theta}}$ using backpropagation.

Strictly speaking, the training process of QNN corresponds to a VQA because an ansatz is trained with training images to minimize a cost function with a classical optimizer.
In other words, the class probability calculation described above is performed iteratively over the input images. 
During inference, the class of each input image is inferred by executing a quantum circuit composed of an encoder and a trained ansatz with $T$ shots.

\subsection{Single flux quantum (SFQ) logic}
\label{background:sfq}

SFQ logic is a digital circuit that utilizes superconductor devices.
At the core of its functionality, SFQ logic processes data through magnetic flux quanta held within superconductor loops. 
These loops incorporate Josephson junctions (JJs), serving pivotal roles as switch mechanisms.
The presence and absence of a single magnetic flux quantum in the loop represent logical `1' and `0', respectively. 
Recently, SFQ has become attractive as a peripheral technology for cryogenic quantum computers because of its high-speed and low-power operation and ability to operate in a cryogenic environment~\cite{holmes2020nisq+,ueno2021qecool,ueno2022qulatis,jokar2022digiq,ravi2023better,byun2022xqsim}.

In this work, we focus on rapid SFQ (RSFQ) technology among the various SFQ circuit technologies to design our architecture. 
Several researchers have developed RSFQ-based microprocessors and demonstrated ultra-high-speed and low-power operations~\cite{FLUX1,SCRAM,Sato_ieeeTr_2017,CORE1}. 
In RSFQ logic, the dynamic energy consumed in a single switching event of a JJ is approximately $10^{-19}$~J.
However, the energy consumed statically due to the voltage and resistance for supplying constant bias current to JJs dominates the total energy consumption.

To eliminate the static power consumption of RSFQ logic, energy-efficient RSFQ (ERSFQ) has been proposed, replacing resistances used for bias current in RSFQ logic with JJs~\cite{mukhanov2011energy}.
In this study, we focus on ERSFQ technology to design our cryogenic architecture.
Given the characteristic of zero static power consumption, ERSFQ technology allows our low-frequency architecture to achieve significantly lower power consumption compared to other technologies, such as cryo-CMOS technology.

%
%
%
%
%
%
%
%


\section{Inter-temperature bandwidth modeling of VQA machines}
\label{analysis}
\subsection{Execution model of VQA}
\label{analysis:execmodel}

\figref{fig:VQA_overview} shows the execution flow of VQA. 
In gate-based quantum computing on NISQ computers, a quantum circuit is executed $T$ times to sample measurement values, where $T$ must be sufficiently large to ensure accurate sampling, typically around $10^6$.
Additionally, VQA iteratively executes a quantum circuit parameterized by $\boldsymbol{\theta}$ using a QPU. 
The parameters $\boldsymbol{\theta}$ are updated by a classical optimizer running on a host computer, based on a cost function calculated from the sampling results.
This hybrid quantum-classical procedure is repeated until the optimal $\boldsymbol{\theta}$ that minimizes the cost function is determined.

The VQA procedure requires two types of inter-temperature communication between a host classical computer and QPU:
\begin{itemize}
    \item \textbf{Instruction transfer:} Quantum circuit information is sent from a host to a QPU.
    \item \textbf{Measurement readout:} Qubit measurement values are sent from a QPU to a host.
\end{itemize}
The instruction transfer communication includes the information of a single quantum circuit, including an ansatz and its parameters.
In the case of VQE, Pauli strings also need to be transferred.
The measurement readout communication transmits the measurement values of $N$ qubits after executing a quantum circuit.
The communication amounts for both instruction transfer and measurement readout increase with $N$.

In this work, we consider two types of VQA execution scenarios: SS and PS executions. 
In the SS execution, a VQA application utilizing $N (\leq M)$ qubits is executed on an $M$-qubit quantum machine with $T$ shots of an $N$-qubit quantum circuit performed sequentially.
In the PS execution, the $T$-shot quantum circuit executions are performed in parallel by fully utilizing all $M$ qubits of the quantum machine.
We denote the number of shots executed simultaneously as $L$, which equals $\lfloor M/N \rfloor$.
This scenario helps bridge the large gap between $N$ and $M$.

\subsection{Baseline system design}
\figref{fig:cryo-vqe}\,(a) shows the baseline system, where we assume an SFQ-based QCI is installed at the 4~K stage. 
The QCI generates SFQ pulses to control and measure the qubits of a QPU, detecting a classical `0' or `1' based on the state of each qubit. 
In addition, we assume that the system can store the information of a quantum circuit used by VQA, such as the ansatz gate sequence and ansatz parameters, within the cryogenic environment during the sampling process. 
Once the quantum circuit information is transmitted to the cryogenic environment, the same quantum circuit can be executed repeatedly for sampling without requiring for additional downlink data.

During VQA execution, inter-temperature communications occur for (1) instruction transfer (downlink) and (2) measurement readout (uplink) as shown in \figref{fig:cryo-vqe}\,(a).
We assume that the baseline system has a sufficient number of communication wires with enough bandwidth to prevent any delays caused by the two types of communication.
The minimum bandwidth needed is referred to as the \textit{required bandwidth} in this paper.
If the required bandwidth is not met, communication delays will increase the execution time of the VQA.

\subsection{Bandwidth for inter-temperature communication}
\label{analysis:bandwidth}

First, we model the required bandwidth for VQE, QAOA, and QML running on the baseline system.
\tabref{tab:summary_baseline_bandwidth} summarizes the modeled bandwidth requirements for each VQA task.

To design a quantum computer system that can execute arbitrary quantum circuits, we need to model the required bandwidth for the worst-case scenario. 
For the worst-case required bandwidth, we assume that the total amount of information for the ansatz parameters $b_{\boldsymbol{\theta}}$ is $O(N)$, and the depth of the ansatz is $O(1)$. 
This implies that the execution time of an ansatz-based quantum circuit, $t_{QC}$, is constant with respect to $N$.
The HEA for VQE shown in \figref{fig:VQA_overview} is a typical example, where the number of parameters $\boldsymbol{\theta} = \lbrace \theta_k^{q, d}\rbrace_{k=1, 2}^{q=1, ..., N, d=1, ..., D+1}$ is $2N(D+1)$, where $D$ is a constant.
In addition, we assume the use of HEA for VQE, QAOA, and QML to simplify the modeling of the required bandwidth for instruction transfer $\BWInst$.

\begingroup
\renewcommand{\arraystretch}{1.28}
\begin{table}[tb]
\caption{Summary of bandwidth models\label{tab:summary_baseline_bandwidth}}
\vspace{-3mm}
\scriptsize
\centering
\begin{tabular}{|c|l|c|} \hline
     & Instruction transfer & Measurement readout              \\ \hline
VQE  & $\BWInst = \BWansatzgatesequence + \BWansatzparams + \bi{\BWpaulistring}$ & \multirow{3}{*}{$\BWMeas = N/t_{QC}$} \\ \cline{1-2}
QAOA & $\BWInst = \BWansatzgatesequence + \bi{\BWansatzparams}$                  &                          \\ \cline{1-2}
QML  & $\BWInst = \BWansatzgatesequence + \bi{\BWansatzparams} +\bi{\BWdata}    $     &                          \\ \hline
\end{tabular}
\begin{tablenotes}
  \item[*] The dominant factor for $\BWInst$ is shown in bold.
\end{tablenotes}
\end{table}
\endgroup


\Figure[t!](topskip=0pt, botskip=0pt, midskip=-2mm){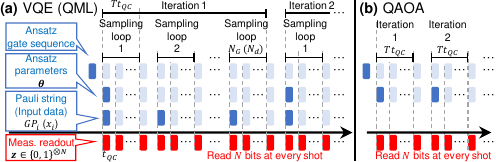}
{Inter-temperature communication timing in the baseline system for VQE (QML) and (b) QAOA with SS scenario.\label{fig:comm-timing}}


\subsubsection{Bandwidth for instruction transfer}
\label{analysis:bandwidth:vqe}

\noindent
\textbf{VQE: }
A quantum circuit used in VQE consists of an ansatz including parameters $\boldsymbol{\theta}$ and a gate sequence for basis transformation used to obtain the expectation value of an inclusive Pauli string $GP$. 
The instruction transfer communication is required to send a gate sequence of the ansatz, ansatz parameters, and inclusive Pauli strings to the cryogenic environment. 
We assume that a gate sequence for basis transformation is appended to the ansatz gate sequence at the 4~K stage, based on an inclusive Pauli string.
We denote the bandwidth required for transmitting each piece of information as $\BWansatzgatesequence$, $\BWansatzparams$, and $\BWpaulistring$, respectively.

An ansatz gate sequence remains constant throughout a VQE execution; therefore, it is transferred only once before the first iteration, as shown in \figref{fig:comm-timing}\,(a).
Transferring it at low speed makes the impact of $\BWansatzgatesequence$ negligible.

Ansatz parameters are transferred once every $N_gT$ shots (\textit{i.e.}, once per iteration), where $N_g$ denotes the number of Pauli groups formed by a grouping method. 
Therefore, $\BWansatzparams$ is modeled as
\begin{align}
    \BWansatzparams = b_{\boldsymbol{\theta}}/N_GTt_{QC}. \label{eq:bw_ap} 
\end{align}
Given that the order of $N_G$ is $O(N^4)$~\cite{Verteletskyi:2020me} or $O(N)$~\cite{Kurita:2023pa}, the order of $\BWansatzparams$ is $O(1/N^3T)$ or $O(1/T)$.

By contrast, an inclusive Pauli string must be transferred once every $T$ shots (\textit{i.e.}, once per sampling loop).
An inclusive Pauli string is a $2N$-bit information that specifies $I/X/Y/Z$ for each qubit. 
Thus, $\BWpaulistring$ is modeled as 
\begin{align}
    \BWpaulistring = 2N/Tt_{QC} \sim O(N/T).  \label{eq:bw_ps}
\end{align}

Based on \eqref{eq:bw_ap} and \eqref{eq:bw_ps}, we show that $\BWpaulistring$ is dominant in the instruction transfer bandwidth for VQE. 
Thus, we model $\BWInst \simeq \BWpaulistring$.

\noindent
\textbf{QAOA: }
In a QAOA execution, an ansatz gate sequence and ansatz parameters must be transferred, and, as in VQE, the impact of $\BWansatzgatesequence$ on the required bandwidth is negligible.
As shown in \figref{fig:comm-timing}\,(b), ansatz parameters are transferred once every $T$ shots (\textit{i.e.}, once per iteration).
Thus, $\BWInst \simeq \BWansatzparams$ is modeled as 
\begin{align}
    \BWInst \simeq b_{\boldsymbol{\theta}}/Tt_{QC} \sim O(N/T). \label{eq:bw_ap_qaoa} 
\end{align}

\noindent
\textbf{QML: }
A quantum circuit used in QML consists of an encoder and an ansatz. 
The instruction transfer communication requires the transmission of a gate sequence of the quantum circuit, ansatz parameters $\boldsymbol{\theta}$, and training data $\boldsymbol{x}$. 
By interpreting an inclusive Pauli string in the context of VQE as training data $x_k$, the same modeling approach applies to the instruction transfer bandwidth for QML.
Here, we assume each training data $x_k$ has a bit width of $b_x$ and that the size of the dataset $\bm{x}$ is $N_d$.
Based on \figref{fig:comm-timing}\,(a), we model $\BWansatzparams$ and the required bandwidth for training data $\BWdata$, which corresponds to $\BWpaulistring$ for VQE, as 
\begin{align}
    \BWansatzparams = \frac{b_{\boldsymbol{\theta}}}{N_dTt_{QC}},\text{ and }
    \BWdata = \frac{b_x}{Tt_{QC}}.  \label{eq:bw_inst_qml}
\end{align}

\subsubsection{Bandwidth for measurement readout}
In the SS scenario, the required bandwidth for measurement readout $\BWMeas$ can be estimated as $ N $ bits per shot for all VQA tasks, as shown in Figs.~\ref{fig:comm-timing}\,(a) and (b).
As a result, we model $\BWMeas$ as 
\begin{align}
  \BWMeas = N / t_{QC}. \label{eq:bw_measure} 
\end{align}

\subsection{Bandwidth bottleneck in sequential-shot (SS) execution}
\label{analysis:design-time-bandwidth}
Based on the bandwidth models presented in \secref{analysis:bandwidth}, we conclude that the inter-temperature bandwidth bottleneck of the baseline system for all VQA tasks is the measurement readout in the SS scenario.
For VQE, the ratio of $\BWInst$ to $\BWMeas$ can be calculated using equations \eqref{eq:bw_ps} and \eqref{eq:bw_measure} as
\begin{align}
    \BWInst/\BWMeas 
    = 2/T \ll 1. 
    \hspace{3mm}
    (\because T\sim10^6) \label{eq:bw_bottleneck_vqe}
\end{align}
Here, $T\sim10^6$ represents the number of shots required to achieve the \textit{chemical accuracy}~\cite{peruzzo2014vqe}, where an energy error from the exact energy is within $1.6 \times 10^{-3}$ Hartree. 
For QAOA, we can also calculate the ratio of $\BWInst$ to $\BWMeas$ using \eqref{eq:bw_ap_qaoa} and \eqref{eq:bw_measure} as 
\begin{align}
    \BWInst/\BWMeas 
    = b_{\boldsymbol{\theta}}/TN \ll 1. 
    \hspace{3mm}
    (\because b_{\boldsymbol{\theta}} \sim O(N))
     \label{eq:bw_bottleneck_qaoa}
\end{align}
For QML, similar to VQE, the ratios of $\BWansatzparams$ and $\BWdata$ to $\BWMeas$ can be calculated using \eqref{eq:bw_inst_qml} and \eqref{eq:bw_measure} as 
\begin{align}
    \frac{\BWansatzparams}{\BWMeas} 
    = \frac{b_{\boldsymbol{\theta}}}{N_dTN} \ll 1, 
    \ \ \text{and} \ \ 
    \frac{\BWdata}{\BWMeas}
    = \frac{b_x}{TN} \ll 1.
    \label{eq:bw_bottleneck_qml}
    \\
    \hspace{3mm} (\because b_{\boldsymbol{\theta}} \sim O(N), \ \ b_x \sim O(N)) \notag
\end{align}

\subsection{Bandwidth bottleneck in parallel-shot (PS) execution}
\label{analysis:multi-thread}

In this subsection, we show that the bandwidth bottleneck in the PS execution is also the measurement readout.
Since $L$ shots of a given VQA task are executed in the PS execution, an ansatz gate sequence, ansatz parameters, inclusive Pauli strings for VQE, and training data for QML are all shared. 
Thus, the bandwidth models for instruction transfer in the PS execution can be obtained by replacing $T$ with $T' = \lceil T/L \rceil$ in the SS execution models.
As a result, the required bandwidth for instruction transfer in the PS execution $\BWInst^\text{(PS)}$ is at most $L$ times that of $\BWInst$ in the SS execution.

On the other hand, since the $N$-bit measurement results need to be obtained and transmitted simultaneously for $L$ shots, the communication amount of the measurement readout at a time increases linearly with $L$.
Thus, the required bandwidth for measurement readout in the PS execution $\BWMeas^\text{(PS)}$ is modeled as $\BWMeas^\text{(PS)} = \BWMeas \times L$.
Consequently, the ratio of $\BWInst^\text{(PS)}$ to $\BWMeas^\text{(PS)}$ can be calculated as 
\begin{align}
    \frac{\BWInst^\text{(PS)}}{\BWMeas^\text{(PS)}} 
    \leq \frac{\BWInst \times L}{\BWMeas \times L} \ll 1.\ (\because \eqref{eq:bw_bottleneck_vqe}\  \text{to}\  \eqref{eq:bw_bottleneck_qml}) \label{eq:bw_bottleneck_multi_thread}
\end{align}

\section{Precomputation at the 4~K stage}
\label{proposal}

\Figure[t!](topskip=0pt, botskip=-2mm, midskip=-2mm){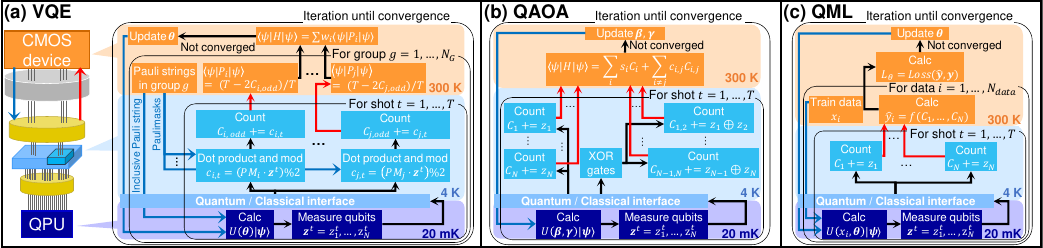}
{Computation overview of (a) VQE, (b) QAOA, and (c) training of QML on C3-VQA architecture.\label{fig:architecture_overview}}


\subsection{Overview of C3-VQA architecture}
As shown in \figref{fig:cryo-vqe}\,(a), the baseline system transfers $N$-qubit measurement results of every shot from the 4~K stage to 300~K environment and calculates expectation values required for VQA in the 300~K environment. 
Thus, the required bandwidth of the baseline system is $\BWMeas = N/t_{QC}$ as discussed in \secref{analysis}, which necessitates a large number of inter-temperature wires. 
As a result, the heat dissipation caused by the inter-temperature wires limits the scalability of the baseline superconducting quantum computer system.

To address the heat dissipation problem by removing the bandwidth bottleneck, we propose the C3-VQA architecture that performs a part of the expectation value calculation for VQA within the cryogenic environment. 
The C3-VQA leverages counter-based processing to exponentially reduce the measurement readout bandwidth relative to the bit width of the counter. 
In addition, the C3-VQA performs the same computations as the baseline and does not affect the solution or convergence of the VQA, although it introduces some execution time overhead compared to the baseline.


The primary goal of the C3-VQA architecture is not bandwidth reduction itself but a reduction in the total heat dissipation in the cryogenic environment, including the passive heat inflow from wires required for bandwidth and the power consumption by the circuits.
Consequently, the C3-VQA architecture should be ultra-low power, and complex calculations for VQA procedures within the cryogenic environment may not be feasible.
More specifically, our cryogenic architecture does not focus on floating-point operations, which lead to high power consumption due to their complex processing, especially in SFQ logic implementations~\cite{nagaoka2023floating}.

In \secref{proposal:motivation}, we show that a part of the expectation value calculations for VQA can be reduced to simple bit operations, which is the focus of the C3-VQA architecture.
In \secref{proposal:counter_based_precomputation}, we explain in detail how the C3-VQA architecture works.
Furthermore, \secref{proposal:overhead} discusses the overhead introduced by C3-VQA, and \secref{proposal:multi-thread} applies the C3-VQA to the PS scenario.

\subsection{Motivation}
\label{proposal:motivation}


In the expectation value calculation of VQE, $C_{i, odd}$ in \eqref{eq:pauli_string_exp} for a given Pauli string $P_i$ is calculated as follows.
We define $PM_i\in \{0, 1\}^N$ as a \textit{Paulimask}, which indicates the non-$I$ positions of $P_i$, and denote $\pz^t \in\{0, 1\}^N$ as $t$-th measurement result of $N$ qubits on the $Z$-basis. 
$c_{i, t} \in \{0, 1\}$ is calculated as
\begin{align}
    c_{i, t} = PM_i \cdot \pz^t \pmod2, \label{eq:c_i_t}
\end{align}
and $C_{i, odd}$ is then calculated as 
\begin{align}
    C_{i, odd} = \sum_t c_{i, t}. \label{eq:C_i_odd} 
\end{align}
For instance, for a Pauli string $P_i = X_0I_1Z_2Y_3$ and $\pz^t = \{1, 0, 1, 0\}$, $c_{i, t}$ is calculated as $c_{i, t} = 1\cdot1 + 0\cdot0 + 1\cdot1 + 1\cdot0 = 0 \pmod 2$.
As a result, we can perform a part of the VQE expectation value calculation using dot product, modulo operations, and counting over $T$ shots.

For QAOA calculation, \eqref{eq:qaoa_cost} and \eqref{eq:qaoa_sampling} can be transformed into
\begin{align}
    \EstQ{H} &= \inv{T}\left(\sum_{i=1}^{N} s_{i} \left(\sum_{t=1}^T {z_{i}^{t}}\right) + \sum_{i \ne j} c_{ij} \left(\sum_{t=1}^T { z_{i}^{t} \oplus z_{j}^{t}}\right) \right).
\end{align}
Here, we define $C_i$ and $C_{ij}$ as follows:
\begin{align}
     C_i = \sum_{t=1}^T {z_{i}^{t}}, \ \ \text{and}\ \ \  
     C_{ij} = \sum_{t=1}^T { z_{i}^{t} \oplus z_{j}^{t}}, \label{eq:qaoa_ci_cij}
\end{align}
respectively.
As a result, a part of the QAOA calculation can be achieved by counting the measurement values of each qubit to compute $C_i$ and performing XOR operations between pairs of $N$ qubits to compute $C_{ij}$ over $T$ shots.


For QML, as derived from \eqref{eq:QML_ff}, we can perform a part of the QML expectation value calculation by simply counting the measurement values of each qubit to compute $C_i$ over $T$ shots.

\figref{fig:architecture_overview} summarizes the execution procedures for VQA on our proposed architecture, as discussed above.
The areas colored in dark blue, light blue, and orange in the diagram represent processes executed in the environments of 20~mK, 4~K, and 300~K, respectively.
The core concept of the C3-VQA architecture, applicable to all VQA variants, involves compressing the information used for expectation value calculation within the cryogenic environment through simple bit operations and counters. 
This approach can significantly reduce the communication from inside to outside the cryostat during a sampling loop.

\subsection{Counting and collecting data}
\label{proposal:counter_based_precomputation}

We assume that the C3-VQA architecture includes $N_C$ counters, each with a bit width of $b$.

\subsubsection{Preprocessing for counters}

As explained in \secref{proposal:motivation}, the expectation value in VQE can be calculated by summing the results of the dot product and modulo calculation between Paulimask $PM_i$ and qubit measurement value $\pz^t$ over a loop for $T$ shots, as shown in equations \eqref{eq:c_i_t} and \eqref{eq:C_i_odd}.
Therefore, the C3-VQA architecture performs the dot product and modulo calculations and counts the results within the cryogenic environment, as illustrated in \figref{fig:architecture_overview}\,(a).
For VQE, our architecture requires $N_C = K$ sets of these components, as shown in \figref{fig:architecture_overview}\,(a).

In the case of QAOA, the procedure in our architecture is simpler compared to that of VQE.
As explained in \secref{proposal:motivation}, the expectation value of QAOA is calculated with summations of each $z_i$ and $z_i \oplus z_j$ over $T$ sampling shots, as shown in \eqref{eq:qaoa_ci_cij}.
Therefore, our architecture requires $N_C = N + \binom{N}{2} = N(N+1)/2$ counters and $\binom{N}{2} = N(N-1)/2$ XOR gates in the cryogenic environment, as shown in \figref{fig:architecture_overview}\,(b).

For QML, the procedure is even simpler than those for the former two.
The loss function of QML is calculated by summing each $z_i$ during a sampling loop, as indicated in equations \eqref{eq:QML_ff} and \eqref{eq:QML_loss_function}.
Hence, our architecture for QML requires $N_C = N$ counters in the cryogenic environment, and they store measurement values of each qubit without any bit operations, as shown in \figref{fig:architecture_overview}\,(c).

\subsubsection{MSB-sending policy of counters}
\label{proposal:counter_based_precomputation:MSB}
The results of the aforementioned bit operations are stored individually in $N_C$ counters, each having a bit width of $b$.
The C3-VQA architecture reduces the required bandwidth for measurement readout by transmitting only the MSBs of counters to the room-temperature environment each time a counter overflows during the sampling loop. 
A $b$-bit counter overflows at most once in every $2^b$ executions of the quantum circuit.
Hence, the bandwidth requirement for sending MSBs in the C3-VQA architecture $\pBWMSB$\footnote{To distinguish the bandwidth associated with the C3-VQA architecture from that of the baseline system, it is denoted with a dot.} is modeled as 
\begin{align}
 \pBWMSB = N_C/2^{b}t_{QC}. \label{eq:bw_msb}
\end{align}
As a result, our architecture achieves an exponential reduction in the measurement readout bandwidth with respect to $b$.

\subsubsection{Collecting non-MSBs of counters}
\label{proposal:counter_based_precomputation:non-MSB}

To calculate the expectation value correctly, it is necessary to account for both the MSBs transmitted during a sampling loop and the non-MSBs stored in each counter at the end of the loop.
Thus, in the C3-VQA architecture, inter-temperature communication is required to send the non-MSBs of each $b$-bit counter to the room-temperature environment after a sampling loop, which introduces execution time overhead. 
In other words, by collecting the non-MSBs stored in each counter, the C3-VQA architecture produces the same VQA solution as the baseline.

We denote the execution time overhead for one sampling loop as $r$ and model the bandwidth for collecting non-MSBs $\pBWnonMSB$ as 
\begin{align}
    \pBWnonMSB = bN_C/rTt_{QC}.
\end{align}
To ensure that $\pBWnonMSB$ remains as small as $\pBWMSB$ in \eqref{eq:bw_msb}, $r$ must satisfy 
\begin{align}
 \pBWnonMSB \leq \pBWMSB 
  \hspace{2mm} \Leftrightarrow \hspace{2mm}
r \geq b2^b/T.     \label{eq:exec_time_overhead_for_nonMSBs}
\end{align}
Hence, our architecture involves a tradeoff between the bandwidth reduction and the execution time overhead; 
the bandwidth reduction is exponential with respect to $b$, while the time overhead also grows exponentially with $b$.

\subsection{Overhead of C3-VQA architecture}
\label{proposal:overhead}

\subsubsection{Collecting non-MSBs in VQE and QML}
\label{proposal:overhead:latency_hiding_technique_for_nonMSB}

\Figure[t!](topskip=0pt, botskip=0pt, midskip=-2mm){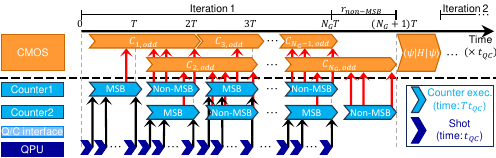}
{Inter-temperature communication in C3-VQA architecture for VQE.\label{fig:MSB_nonMSB_send_timing}}


As discussed in \secref{proposal:counter_based_precomputation:non-MSB}, a naive implementation of our architecture introduces a tradeoff between execution time and bandwidth reduction.
However, we can mitigate the execution time overhead for VQA tasks with the exception of QAOA.

In the VQE procedure, we can start the $G_{i+1}$-th sampling loop before collecting the non-MSBs from each counter for the $G_i$-th sampling loop, as calculations with the results of the $G_i$-th sampling loop are only performed at the end of each iteration.
Hence, the C3-VQA architecture can hide most of the latency associated with non-MSBs by collecting the non-MSBs for the $G_i$-th sampling loop during the $G_{i+1}$-th sampling loop.

To support the overlapped non-MSB collecting policy, two counters are used for each preprocess result.
Each counter is used alternately for MSB sending and non-MSB collection, as shown in \figref{fig:MSB_nonMSB_send_timing}.
Using this technique, the execution time overhead per iteration introduced by our architecture, $r_{\text{non-MSB}}$, is calculated as
\begin{align}
 r_{\text{non-MSB}} = 1/N_G.
\end{align}
Since the order of $N_G$ is $O(N)$ or $O(N^4)$, the execution time overhead of our architecture becomes negligible as $N$ scales.
In this situation, as we collect the non-MSBs of $N_C$ $b$-bit counters within $Tt_{QC}$, we can model the reduced bandwidth for collecting non-MSBs $\pBWnonMSB'$ as
\begin{align}
    \pBWnonMSB' = bN_C/Tt_{QC}.  \label{eq:bw_nonmsb_overlapped}
\end{align}
With the overlapped non-MSB collection, the required measurement readout bandwidth of the C3-VQA, $\pBWMeas$, is 
\begin{align}
    \pBWMeas = \pBWMSB + \pBWnonMSB' = \frac{N_C}{t_{QC}}\left(\frac{1}{2^b}+\frac{b}{T}\right). \label{eq:bw_meas_proposed}
\end{align}

The discussion in this subsubsection applies to QML by replacing `loop over groups' with `loop over data'.
In this case, $r_{\text{non-MSB}}$ is calculated as $1/N_{d}$.
Note that in the QAOA case, where ansatz parameters are updated after every sampling loop, the overlapped non-MSB collecting policy is not applicable, and the required bandwidth and execution time overhead remain as discussed in \secref{proposal:counter_based_precomputation:non-MSB}.

\subsubsection{Paulimasks for VQE precomputation}
\label{proposal:overhead:paulimasks}
Here, we discuss the Paulimasks required for cryogenic computation in the C3-VQA architecture for VQE.
The use of Paulimasks requires the C3-VQA architecture to have a higher instruction transfer bandwidth compared to the baseline system.

In the baseline system, an inclusive Pauli string $GP$ is sent to the cryogenic environment once every sampling loop, as shown in \figref{fig:comm-timing}.
By contrast, in the C3-VQA architecture, up to $K$ separate Paulimasks $PM_i$ are required for each sampling loop to compute $c_{i, t}$ values for all Pauli strings in a group $G$ simultaneously in the cryogenic environment, as these cannot be derived from the $GP$.
Thus, we need to transfer up to $K$ Paulimasks for our system in addition to $GP$, increasing the instruction transfer bandwidth compared to the baseline.

Since each Paulimask consists of $N$ bits, we need to send up to $KN$ bits of information to the cryogenic environment once per sampling loop.
The Pauli string transfer bandwidth for the proposed VQE system $\pBWpaulistring$, required to complete the transmission of the $(G_{i+1})$-th Paulimasks during the $G_i$-th sampling loop, is modeled as
\begin{align}
  \pBWpaulistring = \BWpaulistring + KN/Tt_{QC} = N(K+2)/Tt_{QC}. \label{eq:bw_proposed_ps}
\end{align}
Consequently, the total required bandwidth of the C3-VQA architecture for VQE, denoted as $\pBW{}=\pBWMeas+\pBWpaulistring$, is modeled as 
\begin{align}
  \pBW{} = \left(N(K+2) + K((T/2^b)+b)\right)/Tt_{QC}.  \label{eq:bw_proposed_total}
\end{align}

%

\subsection{C3-VQA architecture for PS execution}
\label{proposal:multi-thread}
While the discussion in this section is based on the SS scenario, it can be applied to the PS scenario by appropriately interpreting the parameters.
Assume that the number of counters required for execution is $N_C'$, and that the C3-VQA architecture has $N_C = LN_C'$ counters.
Since $L$ shots are executed in parallel in PS execution, the total execution time of one sampling loop is $T't_{QC}$, where $T' = \lceil T/L \rceil$.
Therefore, the measurement readout bandwidth of the C3-VQA architecture in the PS scenario can be modeled as
\begin{align}
 \pBWMSB^\text{(PS)} \approx \frac{N_C}{2^{b}t_{QC}},\text{ and }   
 \pBWnonMSB^\text{(PS)} \approx \frac{LN_Cb}{Tt_{QC}}. \label{eq:bw_cacao_multi}
\end{align}
Note that, since the ansatz information for each execution is common in the PS execution, the instruction transfer bandwidth in the PS scenario is the same as in the SS scenario.


\section{Implementation with SFQ circuit}
\label{implementation}
\subsection{Design}
\label{implementation:design}
We design a counter-based co-processor using the well-established RSFQ cell library~\cite{detail_of_cell_library_ADP2}.
The library includes a conventional logic family of RSFQ circuits designed for a niobium nine-layer, 1.0-$\mu$m fabrication process, optimized for operation at 4~K.
\tabref{tab:circuit_configuration} summarizes the SFQ logic gates used in this work with their JJs and bias current (BC) values.


\begin{figure*}[tb]
  \centering
  \includegraphics[width=2.0\columnwidth]{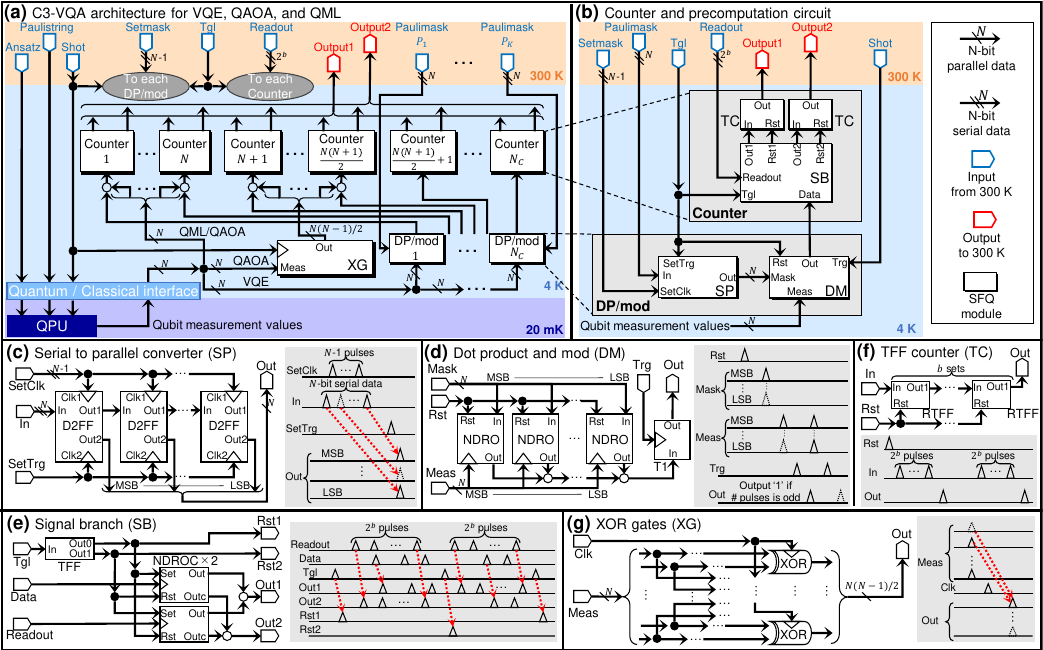}
  \vspace{-2mm}
  \caption{(a) Overview of SFQ implementation of our architecture for VQE, QAOA, and QML. (b) Dot product and modulo arithmetic unit and counter. (c) to (f) Circuit diagram and time chart of each SFQ subcircuit. }
  \label{fig:circuit_diagram}
\end{figure*}

\begin{table*}[tb]
\caption{Detailed configuration of the C3-VQA architecture based on the AIST 10-kA/$\text{cm}^2$ ADP cell library\cite{detail_of_cell_library_ADP2}}
\label{tab:circuit_configuration}
\vspace{-3mm}
\centering
\scalebox{0.9}{

\begin{tabular}{|c|c|c||c|c|c|c|c|c|c|c|} \hline
\multirow{2}{*}{Cell} & \multirow{2}{*}{JJs}  & \multirow{2}{*}{\begin{tabular}[c]{@{}c@{}}BC\\ (mA)\end{tabular}}    &  \multicolumn{2}{c|}{DP/mod}   & \multicolumn{2}{c|}{Counter}& \multirow{2}{*}{\begin{tabular}[c]{@{}c@{}}XG\\ (for QAOA)\end{tabular}}& \multicolumn{3}{c|}{Other}                                \\ \cline{4-7} \cline{9-11} 
             &     &           &  SP          & DM             & SB      & TC ($\times 2$)  &                   & VQE      & QAOA       & QML                     \\ \hline
Splitter     &   3 & 0.30      &  $2(N-1)$    & $N-1$          & 4       & $2(b-1)$           & $3N(N-1)/2$       & $K(N+3)$ & $N+1$      & $N$                     \\
Merger       &   7 & 0.88      &              & $N-1$          & 2       &                  &                   &          & $N(N+1)/2$ & $N$                     \\
RTFF         &  13 & 0.808     &              &                & 1       & $2b$             &                   &          &            &                         \\
T1           &  12 & 0.74      &              & 1              &         &                  &                   &          &            &                         \\
NDRO         &  11 & 1.112     &              & $N$            &         &                  &                   &          &            &                         \\
NDROC2       &  33 & 3.464     &              &                & 2       &                  &                   &          &            &                         \\
D2FF         &  12 & 0.944     &  $N$         &                &         &                  &                   &          &            &                         \\
XOR          &  11 & 1.068     &              &                &         &                  & $N(N-1)/2$        &          &            &                         \\ \hline \hline
Total BC (mA) &     &           & $1.5N-0.6$  & $2.2N-0.44$    & $10.7$  & $2.2b-0.60$      & $0.98N(N-1)$   & $0.3K(N+3)$ &$(N+1)(0.44N+0.3)$  &$1.2N$    \\ \hline
\end{tabular}
}
\vspace{-4mm}
\end{table*}

\figref{fig:circuit_diagram}\,(a) provides an overview of C3-VQA architecture for VQE, QAOA, and QML.
It is assumed that the QPU receives the `Ansatz' and `Paulistring' from the 300~K environment through the quantum/classical interface and executes a quantum circuit triggered by `Shot' inputs.
Additionally, it is assumed that the C3-VQA architecture receives the measurement outcomes of each quantum circuit execution as $N$-bit parallel information from the QPU.
The measurement bitstring is fed into three precomputation units that are selectively activated depending on the target VQA task, and the results are stored in each corresponding `Counter' module.
For VQE, dot product and modulo calculations as described in \eqref{eq:c_i_t} are performed by the `DP/mod' modules, and the results are stored in up to $K$ corresponding Counter modules.
For QML and QAOA, the $N$-bit measurement values are directly stored in $N$ Counters. 
In addition, for QAOA, the results of XOR operations between any pairs of $N$ qubits are stored in additional $N(N-1)/2$ Counters as specified in \eqref{eq:qaoa_ci_cij}.

\figref{fig:circuit_diagram}\,(b) details the connections and input/output of the DP/mod and Counter modules.
As shown in Figs.~\ref{fig:circuit_diagram}(a) and (b), the C3-VQA architecture consists of five submodules, and Figs.~\ref{fig:circuit_diagram}(c) to (g) provide their standard cell-level designs with timing charts (gray area).
\tabref{tab:circuit_configuration} summarizes the standard cells used in this work and provides the detailed configuration of the circuit.
Note that the figures and the table do not account for wiring cells required for SFQ design; the layout of the key building blocks of the C3-VQA in~\secref{implementation:layout} focuses on detailed SFQ circuit design, including such wiring cells.

\noindent
\textbf{Serial to parallel converter (SP)}: 
The SP submodule consists of $N$ D2FF cells and converts the $N$-bit serial input `Paulimask' into parallel data.
Upon receiving the `SetTrg' signal, the $N$-bit parallel signal held in the D2FFs is destructively output to `Out' and transmitted to the DM submodule. 
By receiving the $(G_i+1)$-th Paulimask into the SP while performing the $G_i$-th sampling loop, this submodule supports the bandwidth discussion in \secref{proposal:overhead:paulimasks}.

\noindent
\textbf{Dot product and mod (DM)}: 
The DM submodule comprises $N$ NDRO cells and one T1 cell, performing the dot product and modulo calculations. 
The mask information, converted to parallel by the SP, is held in the NDROs. 
By inputting the $N$-bit parallel `Meas' into the `Clk' input of each NDRO, bitwise AND processing is executed. 
The T1 cell receives the merged outputs of each NDRO, calculating mod 2 of the dot product result.
Upon receiving the `Trg' input, the T1 cell outputs the result of the modulo calculation, namely $c_{i, t}$ in \eqref{eq:c_i_t}, to `Out.'
Note that SP and DM submodules are used only for the VQE task calculation.

\noindent
\textbf{Signal branch (SB)}
The SB submodule distributes inputs from the `In' and `Readout' ports to two distinct outputs, `Out1' and `Out2', ensuring proper allocation. 
Specifically, one output channel is dedicated to the signal received from the `In', while another channel exclusively handles the signal from the `Readout' input. 
Upon receiving the `Tgl' signal, the module switches the ports for the `In' and `Readout' outputs, effectively reversing their roles. 
Simultaneously, this transition triggers the appropriate reset output, `Rst1' or `Rst2'.
The SB plays a pivotal role in alternately utilizing the two connected TFF counters.

\noindent
\textbf{TFF counter (TC)}: 
The TC submodule is a counter consisting of $b$ resettable T-FlipFlops connected in series.
Each pulse input to `In' increments the counter value by one, and upon receiving $2^b$ pulses, the counter overflows, and MSB is sent as a pulse to the `Out' port.
Each Counter module contains two TCs combined with the SB to perform the overlapped non-MSB collection for VQE or QML, as explained in \secref{proposal:overhead:latency_hiding_technique_for_nonMSB}.

\noindent
\textbf{XOR gates (XG)}
The XG submodule consists of $N(N-1)/2$ XOR gates to calculate XORs between any pair of $N$ qubits, and it is used only for QAOA.
XG contains a total of $3N(N-1)/2$ splitters, of which $N(N-1)$ form any pair of $N$ qubits, and $N(N-1)/2$ distribute the `Clk' signal to each XOR gate.
The computed results of the XOR operations are latched in XOR cells and are read out upon receiving the `Clk' signal, which is the `Shot' input that drives the next quantum circuit. 
Thus, there is a one-shot delay in reading out the XOR results compared to the qubit measurements. 
However, given the nature of the C3-VQA architecture, which accumulates the XOR results using counters, the impact of this delay on the expectation value calculation is negligible.


\subsection{Layout}
\label{implementation:layout}
\Figure[t!](topskip=0pt, botskip=0pt, midskip=0pt)[width=0.999\columnwidth]{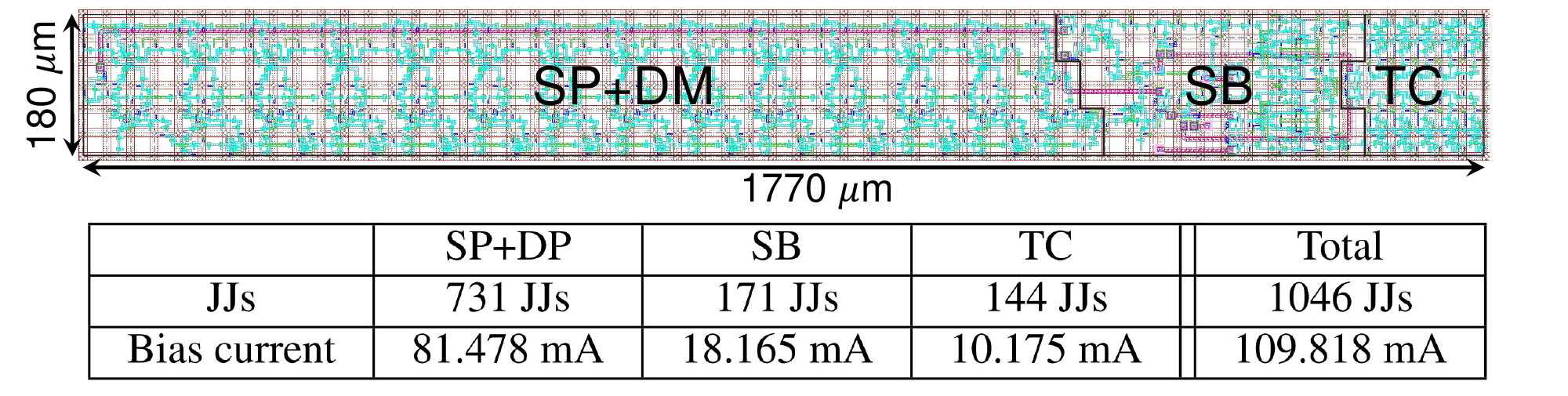}
{Layout of DP/mod ($N=14$) and Counter ($b=5$) modules.\label{fig:chip_layout}}

We used the Josephson simulator (JSIM)~\cite{fang1989josephson}, which is a SPICE-level simulator, to verify the functionality of the proposed architecture.
We then designed the key building blocks of the C3-VQA architecture: the DP/mod and Counter modules shown in \figref{fig:circuit_diagram}\,(b).
For this design, we chose the parameters $N=14$ and $b=5$.
\figref{fig:chip_layout} shows the layout of the design, including the total number of JJs, the area, and the bias current, all based on RSFQ logic.
The designed circuit consists of a total of 1046~JJs, with an area footprint of 0.319 $\text{mm}^2$.

\section{Evaluation}
\label{evaluation}

\subsection{Power estimation of C3-VQA architecture with ERSFQ}
To achieve ultra-low-power consumption for the C3-VQA architecture, we employ the ERSFQ~\cite{mukhanov2011energy}, which eliminates the static power consumption of RSFQ circuits. 
The power consumption of the ERSFQ circuit is estimated using the ERSFQ power model~\cite{mukhanov2011energy}, based on the RSFQ circuit design, as
\begin{align}
P_{\text{ERSFQ}} =2\times \Phi_{0} \times (\text{BC of RSFQ design}) \times (\text{frequency}). \nonumber
\end{align}
Here, we use the magnetic flux quantum $\Phi_0$ of $2.068$~fWb.

The C3-VQA architecture assumes the use of module-level power gating, which reduces the power consumption of unused modules to zero by cutting off the bias current supplied to them. 
Based on the RSFQ design outlined in \tabref{tab:circuit_configuration}, we estimate the power consumption of the essential components of each VQA task execution using the ERSFQ power model.
Note that \tabref{tab:circuit_configuration} does not include wiring cells; therefore, the actual power consumption of a taped-out chip will exceed the estimated power consumption due to wiring overhead.

First, we formulate the execution time of the quantum circuit $t_{QC}$ based on the hardware efficient ansatz for VQE~\cite{kandala2017hardware} to define the circuit frequency $f$ as $f = 1/t_{QC}$. 
For the sake of simple evaluation, we assume that this $t_{QC}$ is common across all VQA tasks.
In practice, the value of $t_{QC}$ is estimated individually for each VQA task, as summarized in the \tabref{tab:quantum_circuit_duration}. 
However, it is important to note that there is no significant difference in the values for the worst-case scenarios requiring the highest bandwidth, and this does not affect the trend of the evaluation in this paper.

We generally denote the time required to apply a quantum gate $ G $ as $t_{G}$.
We assume that each one-qubit (two-qubit) gate operates with the same duration $t_{1Q}$ ($t_{2Q}$) except for reset and measurement operation.
$\ResetT$ and $\MeasureT$ represent the time required for the reset and measurement operations per qubit, respectively.
We assume the maximum gate-level parallelism of one- and two-qubit gates; $M$ one-qubit gates can be applied in parallel to each qubit, and $M/2$ two-qubit gates can be simultaneously applied to each pair of qubits on an $M$-qubit machine.
Then, we model the $t_{QC}$ as in \tabref{tab:quantum_circuit_duration}.
To consider the worst-case scenario for the required bandwidth and circuit power consumption, we suppose the depth $D = 1$.
Using parameters described in references~\cite{jokar2022digiq,opremcak2021measurement}, where $\ResetT = 100$~ns, $t_{1Q} = 10$~ns, $t_{2Q} = 60$~ns, and $\MeasureT = 380$~ns, respectively, $t_{QC}$ is estimated as 
\begin{align}
    t_{QC} \simeq 100 + 2\times (10+10) + 60 \times 2 + 380 = 640_{[\text{ns}]}. \label{eq:t_qc_value}
\end{align}
Based on the $t_{QC}$ value and RSFQ design in \secref{implementation}, we estimate the power consumption of C3-VQA for each VQA task when ERSFQ is applied in the following parts.

\begingroup
\renewcommand{\arraystretch}{1.3}
\begin{table}[tb]
\centering
\scriptsize
\caption{Quantum circuit duration for each VQA task\label{tab:quantum_circuit_duration}}
\vspace{-3mm}
\begin{tabular}{|c|l|} \hline
VQA task  & $t_{QC}    $                                            \\ \hline\hline
VQE\cite{kandala2017hardware} & $\ResetT+(D+1)(\ryT+\rzT)+2\czT+\MeasureT$              \\ \hline
QAOA\cite{farhi2014qaoa}      & $\ResetT+\InitT+p(\ryT+2(2\cxT+\rzT)+\ryT) +\MeasureT $ \\ \hline
QML                           & $\ResetT+\encodeT+\ryT + 2\czT+ \rxT +\MeasureT$       \\ \hline
\end{tabular}
\vspace{-4mm}
\end{table}
\endgroup

\subsubsection{VQE}
We assume the C3-VQA architecture includes $N$ qubits, $N_C$ Counter modules, and $N_C$ DP/mod modules, with each Counter having a bit width of $b$.
According to \tabref{tab:circuit_configuration}, the essential modules of our architecture for the VQE task include $N_C$ DP/mod modules, $N_C$ Counter modules, and Other components specific to VQE.
We assume the operating frequency of the entire circuit $f = 1/t_{QC}$ as 1.56~MHz except for the SP submodule.
That of the SP submodule is assumed to be $fN/T$ due to its $N$-bit serial inputs and a processing speed of $1/T$.
The power consumption for VQE with ERSFQ can be estimated as a function of $N, N_C, b$, and $T$:
\begin{flalign}
  &P_{\text{VQE}}(N, N_C, b, T)&  \label{eq:vqe_power} \\ 
  &\quad= 2\Phi_{0}f\times N_C(N(1.5N-0.6)/T + (2.2N-0.44)+10.7& \nonumber \\
  &\qquad \qquad \qquad \quad +(2.2b-0.6)+0.3(N+3))& \nonumber \\ 
  &\quad\approx 6.5N_C\left(N(1.5N-0.6)/T + 2.5N +10.6+2.2b\right)_{[\text{pW}]}.& \nonumber
\end{flalign}

\subsubsection{QAOA}

According to \figref{fig:architecture_overview}\,(a) and \tabref{tab:circuit_configuration}, the essential modules are $N(N+1)/2$ Counter modules along with Other components for QAOA.
Similar to VQE, the power consumption for QAOA with ERSFQ can be estimated as a function of $N$ and $b$:
\begin{flalign}
  &P_{\text{QAOA}}(N, b)&  \label{eq:qaoa_power} \\ 
  &\quad= 2\Phi_{0}f\times ((10.7+2.2b-0.6)\times N(N+1)/2 & \nonumber\\
  &\qquad \qquad \quad \quad + 0.98N(N-1)+(N+1)(0.44N+0.3))& \nonumber \\ 
  &\quad\approx 6.5\left((1.1b+6.5)N^2 + (1.1b+4.9)N +0.3\right)_{[\text{pW}]}.& \nonumber
\end{flalign}

\subsubsection{QML}
The essential modules are $N$ Counter modules along with Other components for QML.
As with the VQE and QAOA, the power consumption for QML can be estimated as a function of $N$ and $b$:
\begin{align}
  P_{\text{QML}}(N, b)  &= 2\Phi_{0}f\times (10.7+2.2b-0.6+1.2)\times N \nonumber \\
  &\approx 6.5N(2.2b+11.3)_{[\text{pW}]}. \label{eq:qml_power}
\end{align}

\subsection{Scalability to the number of qubits}
\label{evaluation:scalability}


\begin{figure*}[tb]
  \centering
  \includegraphics[width=1.95\columnwidth]{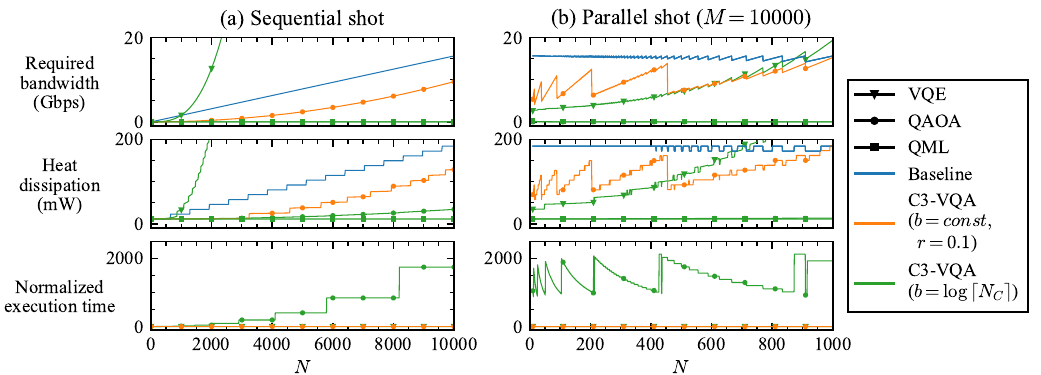}
  \caption{Required bandwidth (top) and power dissipation (middle, solid line) of the baseline and C3-VQA architecture are shown for (a) SS and (b) PS scenarios. The bottom plots show the execution time of C3-VQA normalized to that of the baseline. }
  \label{fig:eval_single_multi}
  \vspace{-5mm}
\end{figure*}

We evaluate the total heat dissipation in the cryogenic environment with the C3-VQA architecture.
We assume that the baseline and proposed systems use stainless steel coaxial cables (UT085 SS-SS) between the 4~K and upper stages, with a passive heat inflow of 1.0~mW as reported in~\cite[Table 2]{krinner2019cryosetup}. 
In addition to the heat inflow, we consider the power consumption of the wire peripherals.
We assume the amplifier, which is the main source of peripheral power consumption, to be the LNF-LNC4\_8C~\cite{krinner2019cryosetup}, with a power consumption of 10.5~mW per wire, as specified in its datasheet. 
We assume the bandwidth per wire is 1~Gbps, and the system is considered to have a sufficient number of wires to exceed its required bandwidth.
\tabref{tab:cable_counter_configuration} summarizes the configuration of the wires and the C3-VQA architecture.

Note that the primary focus of this paper is the relationship between inter-temperature communication bandwidth and the associated heat dissipation in the superconducting quantum computer systems. 
Therefore, the subsequent evaluations focus on the heat inflow via wires and the power consumption of wire peripherals and the C3-VQA architecture, while excluding QCIs from the evaluation.
In addition, while our evaluation utilizes a specific coaxial cable and peripheral, it should be noted that the C3-VQA is expected to perform well across a variety of system designs because of its general advantages.

\begingroup
\renewcommand{\arraystretch}{1.5}
\begin{table}[tb]
\centering
\scriptsize
\caption{Configuration of cables and C3-VQA architecture}\label{tab:cable_counter_configuration}
\vspace{-3mm}
\scalebox{0.9}{

\begin{tabular}{|ll|l|l|} \hline
\multicolumn{2}{|l|}{}                                  & Heat dissipation per unit                                        & Bandwidth \\ \hline
\multicolumn{2}{|c|}{Coaxial cable}                     & \begin{tabular}[c]{@{}l@{}}Heat inflow: 1.0 mW\cite{krinner2019cryosetup}\\ Peripherals: 10.5 mW\cite{krinner2019cryosetup}\end{tabular}     & 1~Gbps per one cable                                                                   \\ \hline
\multicolumn{1}{|c|}{\multirow{2}{*}{C3-VQA}} & SS      & \begin{tabular}[c]{@{}l@{}}VQE: $P_{\text{VQE}}(N, K, b, T)$  \\ QAOA: $P_{\text{QAOA}}(N, b)$\\ QML: $P_{\text{QML}}(N, b)$\end{tabular}    & \begin{tabular}[c]{@{}l@{}}$\pBWMSB+\pBWnonMSB+\pBWpaulistring$\\ $\pBWMSB$\\ $\pBWMSB+\pBWnonMSB$\end{tabular} \\ \cline{2-4}
\multicolumn{1}{|c|}{}                        & PS      & \begin{tabular}[c]{@{}l@{}}VQE: $P_{\text{VQE}}(N, LK, b, T')$\\ QAOA: $P_{\text{QAOA}}(N, b)$\\ QML: $P_{\text{QML}}(N, b)$\end{tabular}    & \begin{tabular}[c]{@{}l@{}}$\pBWMSB^\text{(PS)}+\pBWnonMSB^\text{(PS)}+\pBWpaulistring  $\\ $\pBWMSB^\text{(PS)} $\\ $\pBWMSB^\text{(PS)}+\pBWnonMSB^\text{(PS)}$\end{tabular} \\ \hline
\end{tabular}
}
\vspace{-5mm}
\end{table}
\endgroup

\subsubsection{SS scenario}
First, we present the scalability results for the SS execution scenario, where a single VQA program utilizing $N$ qubits is executed.
Here, we assume $K = N_G = N^2$ in the case of VQE, and the number of shots is set to $T = 10^6$.

\figref{fig:eval_single_multi}\,(a) shows the required bandwidth and total heat dissipation for the baseline and proposed systems during the SS execution in the top and middle plots, respectively. 
The color of the lines distinguishes between the baseline system and the two configurations of the C3-VQA, while the markers on each line represent the VQA tasks.
Note that the blue lines for the baseline system have no marker as the required bandwidth and total heat dissipation of the baseline are the same for all VQA tasks.

The total heat dissipation of the baseline system (blue solid line) increases in a stepwise manner, indicating an increase in the number of wires corresponding to the bandwidth requirement, which scales linearly with $N$.
The C3-VQA with counter bit width $b = \log{\lceil N_C \rceil}$ has constant bandwidth for QAOA and QML (the green lines with circles or squares). 
However, the bottom plot (green line with circles) shows that the C3-VQA is not practical for QAOA due to its large execution time overhead, as discussed in \secref{proposal:counter_based_precomputation:non-MSB}.
Note that the execution time overhead for the VQE and QML is negligible as discussed in \secref{proposal:overhead:latency_hiding_technique_for_nonMSB}.
As a result, when $N$ is 10,000, the C3-VQA reduces total heat dissipation by 94\% compared to the baseline for QML, with negligible execution time overhead. 

For the QAOA case, we present another result with C3-VQA with a constant counter bit width $b$ (orange lines).
We allow a 10\% execution time overhead $r$, and $b$ is determined by \eqref{eq:exec_time_overhead_for_nonMSBs}.
The C3-VQA with constant $b$ reduces required bandwidth and heat dissipation for the QAOA task with $N=10,000$ by 39\% and 30\%, respectively.

For VQE (lines with triangles), the required bandwidth of C3-VQA increases more rapidly than that of the baseline because $\pBWpaulistring$ in \eqref{eq:bw_proposed_ps} becomes dominant as $K$ increases with $N$.
Additionally, the power consumption of the C3-VQA scales with $O(N^3)$ as formulated in \eqref{eq:vqe_power}, resulting in advantages over the baseline within a limited range of $N$.

\subsubsection{PS scenario}
Next, we consider the PS execution of an $N$-qubit VQA task on a $10,000$-qubit machine.
Let $L = \lfloor 10,000/N \rfloor$ represent the number of shots executed simultaneously, with the group size for a given VQE program denoted as $K = N^2$.
We assume the total number of shots $T = 10^6$; when executing $L$ shots simultaneously, the execution time of one sampling loop is given by $T't_{QC}$, where $T' = T/L$.
Other parameters are set in the same manner as those for the SS scenario.

\figref{fig:eval_single_multi}\,(b) shows the results for the PS scenario.
For all tasks, both the total heat dissipation and required bandwidth of the baseline system remain constant with respect to $N$, except for differences due to the remainder of $M/N$ (blue lines).
For QAOA and QML, similar to the SS scenario, the C3-VQA with $b=\log{\lceil N_C \rceil}$ maintains the required bandwidth as a constant relative to $N$.
However, the execution time overhead for QAOA is significant.

For VQE, the required bandwidth and total heat dissipation of the proposed system increase mildly as $N$ scales. 
The advantage of the proposed system in reducing heat dissipation compared to the baseline is more significant for smaller $N$.
For instance, the proposed system reduces heat dissipation by 81\% for $N = 20$ and by 74\% for $N = 100$. 


\subsubsection{VQE case study for quantum chemistry}

\Figure[t!](topskip=0pt, botskip=0pt, midskip=0pt){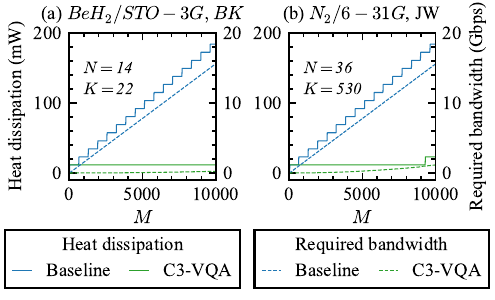}
{VQE case studies for two quantum chemistry problems~\cite{Verteletskyi:2020me}.\label{fig:eval_case_study}}


To validate the effectiveness of the C3-VQA architecture in practical problem settings, we estimate heat dissipation for VQE tasks in quantum chemistry.
We assume the ``$BeH_2/STO-3G$'' (``$N_2/6-31G$'') problem with Bravyi–Kitaev (Jordan-Wigner) transformed Hamiltonian, as presented in~\cite[Table II]{Verteletskyi:2020me}, representing the smallest (largest) size problem in the reference.

Figs.~\ref{fig:eval_case_study}\,(a) and (b) show the bandwidths and heat dissipation for the baseline and C3-VQA in the PS scenario.
While the baseline system shows a similar trend to the SS case, the proposed system maintains lower bandwidth and heat dissipation, even as $M$ scales.
Specifically, for $M=10,000$, as shown in \figref{fig:eval_case_study}\,(b), where $L = 277$ shots for $N_2/6-31G$ problem are executed simultaneously, the proposed system reduces the required bandwidth by 93\%, resulting in a heat dissipation reduction of 87\% compared to the baseline.

\section{Summary and future prospect}
\label{sec:conclusion}
In this paper, we have modeled the execution time and inter-temperature communication of VQAs in cryogenic quantum computers and proposed the C3-VQA, a cryogenic counter-based co-processor, to enhance the design scalability of cryogenic quantum computer systems.
The evaluation shows that the proposed system successfully reduced the required bandwidth with small power consumption, resulting in a significant reduction in total heat dissipation in the PS scenario on a large-scale cryogenic NISQ machine. 
As a result, the C3-VQA architecture, which performs well in the PS scenario, helps bridge the large gap between NISQ and FTQC and promotes the continuous development of quantum computers.

While this paper has primarily focused on addressing bandwidth issues in NISQ machines, we expect that the versatility of the C3-VQA architecture will also be effective in addressing the bandwidth challenges in FTQC, as been highlighted in several studies~\cite{tannu2017taming,ravi2023better,smith2023local}.
For example, the counter-based approach of the C3-VQA is expected to be used to compress syndrome values for QEC decoding.
The C3-VQA architecture not only addresses current challenges in NISQ devices but also lays the groundwork for scalable quantum computing in the FTQC era.


\bibliographystyle{IEEEtranS}
\bibliography{refs}

\EOD

\end{document}